\documentclass[amsmath,longbibliography,preprint]{revtex4-2}  
\usepackage{bbm}
\usepackage{mathrsfs}
\usepackage{amsfonts}
\usepackage[colorlinks=true,citecolor=blue,anchorcolor=blue]{hyperref}
\usepackage{graphicx,epstopdf}
\usepackage{subfigure}
\usepackage{epsfig}
\usepackage{dcolumn}
\usepackage{bm}
\usepackage{color}
\usepackage{natbib}
\usepackage{amssymb}
\usepackage{xcolor}
\usepackage{braket}
\usepackage{soul}

\usepackage{float}
\usepackage{bm}
\usepackage[ruled,vlined]{algorithm2e}
\usepackage{orcidlink}

\usepackage[english]{babel}

\renewcommand\thesection{\arabic{section}}

\graphicspath{{pict/}{}}

\usepackage[raggedleft]{titlesec}
\titleformat{\section}{\bf\large}{\thesection.\,}{0.24em}{}
\titlespacing{\section}{0cm}{0.5cm}{0cm}

\def\ket#1{\left|{#1}\right\rangle}
\def\braket#1#2{\left\langle{{#1}}\mathrel{\left|{\vphantom{{#1}{#2}}}\right.\kern-\nulldelimiterspace}{{#2}}\right\rangle}

\begin{document}


\title{Adaptive Robust High-Precision Atomic Gravimetry}

\author{Jinye Wei$^{1,2}$}

\author{Jiahao Huang$^{1,2}$}

\author{Chaohong Lee$^{2,3}$}
\altaffiliation{Email: chleecn@szu.edu.cn, chleecn@gmail.com}

\affiliation{$^{1}$Laboratory of Quantum Engineering and Quantum Metrology, School of Physics and Astronomy, Sun Yat-Sen University (Zhuhai Campus), Zhuhai 519082, China}

\affiliation{$^{2}$Institute of Quantum Precision Measurement, State Key Laboratory of Radio Frequency Heterogeneous Integration, College of Physics and Optoelectronic Engineering, Shenzhen University, Shenzhen 518060, China}

\affiliation{$^{3}$Quantum Science Center of Guangdong-Hong Kong-Macao Greater Bay Area (Guangdong), Shenzhen 518045, China}


\begin{abstract}
Atomic gravimeters are the most accurate sensors for measuring gravity, yet a significant challenge lies in achieving high precision while also maintaining high dynamic range and robustness.
Here, we develop a protocol for achieving robust high-precision atomic gravimetry based upon adaptive Bayesian quantum estimation.
Our protocol incorporates a sequence of interferometry measurements taken with short to long interrogation times and offers several crucial advantages. 
Firstly, it enables a high dynamic range without the need to scan multiple fringes for pre-estimation, making it more efficient than the conventional frequentist method. 
Secondly, it improves robustness against noise, allowing for a significant improvement in measurement precision in noisy environments. 
The enhancement can be more than $5$ times for a transportable gravimeter [Sci. Adv. \textbf{5}, eaax0800 (2019)] and up to an order of magnitude for a state-of-the-art fountain gravimeter [Phys. Rev. A \textbf{88}, 043610 (2013)]. 
Notably, by optimizing the interferometry sequence, our approach can improve the scaling of the measurement precision ($\Delta g_{est}$) versus the total interrogation time ($\tilde{T}$) to $\Delta g_{est} \propto \tilde{T}^{-2}$ or even better, in contrast to the conventional one $\Delta g_{est} \propto \tilde{T}^{-0.5}$.
Our approach offers superior precision, increased dynamic range, and enhanced robustness, making it highly promising for a range of practical sensing applications. 

\end{abstract}

\maketitle


\newpage
\section{Introduction\label{Sec1}}
\begin{figure*}[htbp]
\includegraphics[width=1.0\linewidth,scale=1.00]{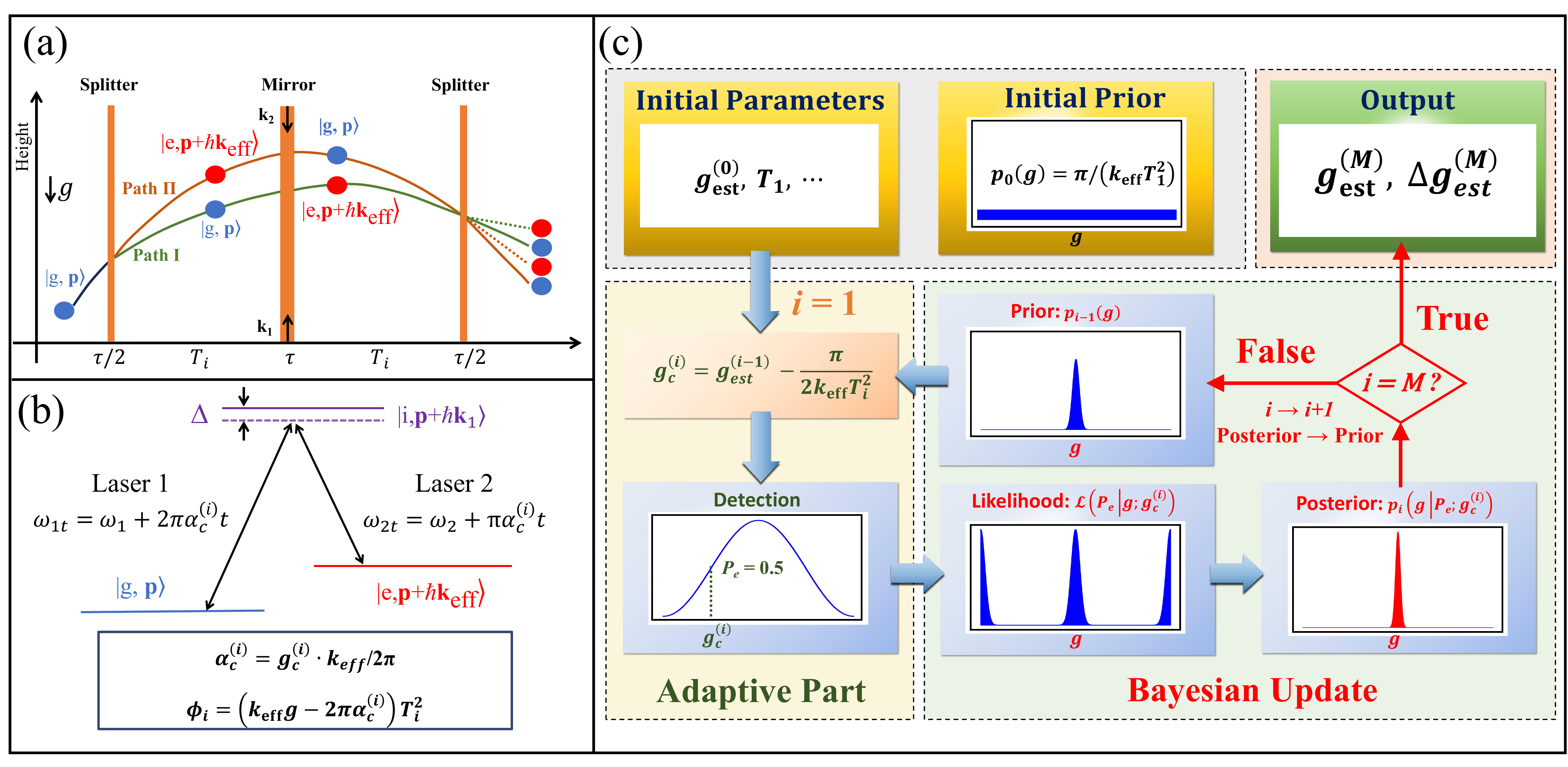}%
\caption{\label{BayesianGravimeter}   
(a) Bayesian atomic gravimetry incorporates a sequence of measurements taken with short to long $T_i$. Here, $T_i$ is the separation time between neighboring pulses in $i^{th}$ iteration. 
(b) Two Raman lasers couple two states $\ket{g, \textbf{p}}$ and $\ket{e, \textbf{p}+ \hbar\textbf{k}_{\rm{eff}}}$ with frequencies $\omega_{1t}=\omega_1+2\pi\alpha_c^{(i)} t$ and $\omega_{2t}=\omega_2+\pi\alpha_c^{(i)} t$. 
(c) Schematic of our adaptive Bayesian gravity estimation. 
Given a prior distribution $p_{i-1}$ and the likelihood function, the posterior distribution $p_{i}$ is updated through the Bayes' formula. 
The estimation value of $g_{est}^{(i)}$ and its standard deviation $\Delta g_{est}^{(i)}$ can be computed with $p_{i}$. Then, $g_c^{(i)}$ is adaptively given and then used to determine $\alpha_c^{(i)}$. Generally, the initial prior $p_0$ is an evenly distributed function.}
\end{figure*}

\maketitle
\noindent
Atom interferometry has become a prominent inertial-sensing technique, in particular, gravimetry.
Atomic gravimetry~\cite{peters1999measurement,peters2001high,cronin2009optics,poli2011precision,altin2013precision,kritsotakis2018optimal,bongs2019taking,szigeti2020high,li2023continuous} plays a vital role in inertial navigation~\cite{jekeli2005navigation,han2016improved,cheiney2018navigation,zhu2024miniaturized}, geophysics ~\cite{lambert2001new,park2005earth,montagner2016prompt,chen2006satellite}, and fundamental research~\cite{tarallo2014test,rosi2015measurement,kovachy2015quantum,duan2016test,asenbaum2017phase,asenbaum2020atom,graham2013new,zhang2023ultrahigh}.
It is important to improve precision, stability and efficiency of atomic gravimeters~\cite{dimopoulos2007testing,lautier2014hybridizing,altschul2015quantum,wu2019gravity,guo2023vibration,panda2023atomic}. 
The gravitational acceleration $g$ is usually determined using Mach-Zehnder interferometry. 
This method utilizes a $\pi/2-\pi-\pi/2$ Raman pulse sequence to coherently split, reflect, and recombine the atomic wave packets, as depicted in figure~\ref{BayesianGravimeter}(a).
The beam-splitters and mirrors are achieved by state-changing Raman transitions, which are achieved with two counter-propagating laser pulses coherently coupling two internal states $\ket{g}$ and $\ket{e}$. 
The first $\pi/2$ pulse transfers the atoms from $\ket{g, \textbf{p}}$ ($\textbf{p}$ labelling the momentum) into an equal superposition of states $\ket{g, \textbf{p}}$ and $\ket{e, \textbf{p}+ \hbar\textbf{k}_{\rm{eff}}}$, obtaining an effective momentum of $\hbar \textbf{k}_{\rm{eff}}=\hbar k_{\rm{eff}} \hat {\textbf{z}}=\hbar (\textbf{k}_1-\textbf{k}_2)=2\hbar k \hat {\textbf{z}}$, where $\textbf{k}_1=-\textbf{k}_2=k \hat {\textbf{z}}$ are the wavevectors of two lasers along the gravity direction $\hat {\textbf{z}}$.
The $\pi$ pulse then completely exchanges the two atomic states, and the second $\pi/2$ pulse recombine the atoms. 
Finally, one can perform the population detection to determine $g$ according to $\mathcal{L}_{u} = \frac{1}{2}\left[1 + (-1)^u \cos \Phi\right]$, 
where $\Phi=k_{\rm{eff}}g T^2$ is the accumulated phase, $T$ represents half of the interrogation time, and $u = 0~\textrm{or}~1$ stands for the atoms occupying $\ket{g, \textbf{p}}$ or $\ket{e, \textbf{p}+ \hbar\textbf{k}_{\rm{eff}}}$, respectively~\cite{baryshev2015application}.  

However, it is a great challenge to simultaneously achieve high precision and high dynamic range. 
On the one hand, although long interrogation time leads to high precision, it will bring phase ambiguity leading to inaccuracy.
Typically, one can tune the laser frequency with a chirp rate $\alpha$ to introduce an auxiliary phase.
Consequently, the accumulated phase becomes $\Phi=(k_{\rm{eff}}g-2\pi \alpha) T^2$.
By scanning at least three fringes with different interrogation times, one can obtain a common $\alpha_0$ and uniquely determine $g=2\pi \alpha_0 / {k}_{\rm{eff}}$~\cite{carraz2009compact,zhou2011measurement,abend2016atom,freier2016mobile,samuel2022using}.
This pre-estimation process requests more measurement times in experiments. 
On the other hand, the dynamic range becomes narrower for longer interrogation times and the corresponding robustness decreases.  
While a short interrogation time offers higher dynamic range but lower precision and a long interrogation time offers higher precision but lower dynamic range, can we combine measurements taken with  short and long interrogation times to simultaneously achieve high precision and high dynamic range?

Bayesian quantum estimation may combine measurements taken with different interrogation times by updating the probability distribution with Bayes' theorem. 
It simultaneously provides excellent robustness, high precision, and wide dynamic range.
This intriguing feature attracts increasing attentions in quantum phase estimation~\cite{berry2000optimal,paesani2017experimental,wang2017experimental,lumino2018experimental,nolan2021machine,qiu2022efficient,PhysRevA.76.033613, Gebhart2023} and its applications in quantum magnetometers~\cite{ruster2017entanglement,santagati2019magnetic,puebla2021versatile,nusran2012high,bonato2016optimized, PhysRevA.106.052603, Craigie2021} and atomic clocks~\cite{PhysRevApplied.22.044058,cao2024multiqubit}.
It has been demonstrated that the precision scaling with respect to the total interrogation time may surpass the standard quantum limit (SQL)~\cite{said2011nanoscale} or even approach the Heisenberg scaling~\cite{higgins2007entanglement,wiebe2016efficient,degen2017quantum}. 
Unlike typical quantum phase estimation, where the accumulated phase has a linear dependence on interrogation time $\Phi \propto T$, in atomic gravimetry, the accumulated phase exhibits a quadratic dependence on interrogation time $\Phi \propto T^2$.
Up to date, it is still unknown that how Bayesian quantum estimation can be utilized to improve the performance of atomic gravimetry.

In this article, we propose an adaptive Bayesian gravity estimation (BGE) protocol to achieve robust and high-precision atomic gravimetry. 
On one hand, our protocol enables a high dynamic range without the need to scan at least three fringes for pre-estimation.
On the other hand, our protocol enhances resilience to noise, enabling over a tenfold improvement in gravity measurement precision in practical noisy conditions. 
Moreover, by optimizing the interferometry sequence of growing interrogation times, the scaling of the measurement precision $\Delta g$ versus the total interrogation time $\tilde{T}$ can be improved from $\Delta g\propto \tilde{T}^{-0.5}$ to $\Delta g\propto \tilde{T}^{-2}$ or even $\Delta g\propto \tilde{T}^{-2.25}$. 
Our BGE protocol can be readily applied to enhance the performance of all existing interferometry-based atomic gravimeters.

\section{Bayesian gravity estimation}

\noindent

Below we show how to employ Bayesian quantum estimation to determine the gravity $g$ without the need to scan three different fringes. 
The key idea is using the measurement taken with short interrogation time (which has high dynamic range) for rough estimation and then utilizing the measurement taken with long interrogation time for high-precision estimation. 
We use Bayesian update to combine the measurements taken with different interrogation times $T_i$, where $i=1,2,...,M$ is the iteration index and $M$ denotes the total steps. 
To perform the adaptive BGE, we introduce an auxiliary parameter $g_c^{(i)} = 2\pi \alpha_c^{(i)}/{k}_{\rm{eff}}$, which is realized by varying the two laser frequencies according to  $\omega_{1t}=\omega_1+2\pi\alpha_c^{(i)} t$ and $\omega_{2t}=\omega_2+\pi\alpha_c^{(i)} t$, see figure~\ref{BayesianGravimeter}(b).
Therefore, the corresponding likelihood function reads
\begin{equation}
    \mathcal{L}_{u} = \frac{1}{2}\left\{ 1 + (-1)^u \cos [(g-g_c^{(i)}) k_{\rm{eff}} T_i^2]\right\}.
\label{eq:singlelikelihood2}
\end{equation} 

Different from conventional schemes, the adaptive BGE can search the optimal $g_c^{(i)}$ for each iteration.
This ensures that the atomic gravimeter operates under the optimal condition with sharpest slope, see figure~\ref{BayesianGravimeter}(c). 
In the beginning, if there is no prior knowledge of $g$, the first prior distribution can be set as an uniformly distributed function.
For an ensemble of $R$ atoms, the likelihood function can be approximated as a Gaussian function~\cite{dinani2019bayesian} $\mathcal{L}(P_e|g; g_c^{(i)}) = \frac{1}{\sqrt{2\pi} \sigma}\exp\left[-\frac{(P_e - \mathcal{L}(1|g;g_c^{(i)}))^2}{2\sigma^2}\right]$,
with $\sigma^2 \approx P_e(1-P_e)/R$ and $P_e$ denoting the probability of atoms occupying the state $\ket{e, \textbf{p}+ \hbar\textbf{k}_{\rm{eff}}}$. 

Analytically, since $g_c^{(i)}$ satisfies the adaptive relation $g_c^{(i)}=g_{est}^{(i-1)}-\pi/(2k_{\rm{eff}}T_i^2)$, we set $\Phi \approx k_{\rm{eff}}(g-g_c^{(i)})T_i^2$ with $\Phi=\arccos(1-2P_e)$ and expand the likelihood around $\Phi = \pi/2$ using Taylor's formula, we can finally obtain 
\begin{equation}
\begin{aligned}
\mathcal{L}(P_e|g; g_c^{(i)}) &\approx \frac{1}{\sqrt{2\pi} \sigma}\exp\left[-\frac{(P_e - \frac{1}{2}(1-(\cos\Phi - \sin\Phi(k_{\rm{eff}}(g-g_c^{(i)})T_i^2-\Phi))))^2}{2\sigma^2}\right] \\
&=\frac{1}{\sqrt{2\pi} \sigma}\exp\left[-\frac{(g-(g_c^{(i)}+A))^2}{2\sigma_i^2}\right], 
\label{eq:likelihoodG}
\end{aligned}
\end{equation}
with
\begin{equation}
    A = [(2P_e - 1 - \cos\Phi)/\sin\Phi + \Phi]/(k_{\rm{eff}}T_i^2),
\end{equation}
and
\begin{equation}
    \sigma_i = 2\sigma/(|\sin\Phi| k_{\rm{eff}}T_i^2).
\end{equation}
Thus the likelihood function of the atomic ensemble is also a Gaussian function versus $g$ with the standard deviation $\sigma_i = \frac{1}{\sqrt{R}\textbf{k}_{\rm{eff}}T_i^{2}}$, which can be applied for the derivation of gravity precision scaling below. 
The information of $g$ can be updated through Bayes' theorem, resulting in a posterior distribution,
\begin{equation}
    p_i (g|P_e; g_c^{(i)})=\mathcal{N}\mathcal{L}(P_e|g; g_c^{(i)}) p_{i-1} (g)
\label{eq:Bayes}
\end{equation}
where $\mathcal{N}$ is the normalization factor, $p_i$ and $p_{i-1}$ denote the posterior and prior distributions, respectively. 
According to the posterior distribution, the estimation of $g$ can be given as $g_{est}^{(i)} = \int g p_i (g|P_e; g_c) dg$ with a standard deviation $\Delta g_{est}^{(i)} = \sqrt{\int g^2 p_ib (g|P_e; g_c)dg - (g_{est}^{(i)})^2}$.
 
\begin{figure*}[htp]
\includegraphics[width=1.0\linewidth,scale=1.00]{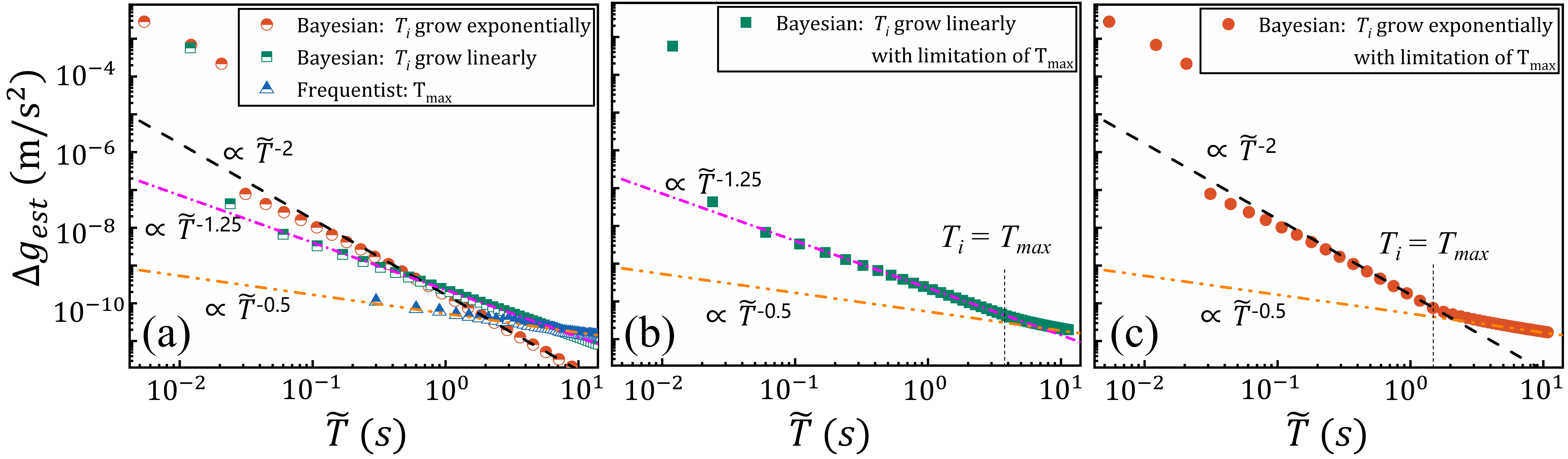}
\caption{\label{Scaling} Precision scaling with respect to total interrogation time. 
(a) $T_{i}$ grow linearly ($b = T_1 = 12\times 10^{-3}$~s) and exponentially ($a = 1.25$, $T_1 = 5.4 \times 10^{-3}$~s), respectively. 
(b) $T_{i}$ grows linearly until it reaches $T_{max}$. 
The precision follow $\tilde{T}^{-1.25}$ until $T_i$ attains $T_{max}$, then eventually converge to $\tilde{T}^{-0.5}$ as $T_i$ remains $T_{max}$ in the following. 
(c) $T_{i}$ grows exponentially until it reaches $T_{max}$. 
The precision follows $\tilde{T}^{-2}$ until $T_i$ attains $T_{max}$, then eventually converges to $\tilde{T}^{-0.5}$ as $T_i$ remains $T_{max}$ in the following.  
Here, $k_{\rm{eff}} = 1.61\times 10^{7}$~rad/m, $T_{max} = 0.3$~s and $R = 5 \times 10^7$. The total measurement times for (b) and (c) is $M=50$. Here, $k_{\rm{eff}} = 1.61\times 10^{7}$~rad/m, $T_{max} = 0.3$~s and $R = 5 \times 10^7$ are chosen based on typical atomic fountain experiments.}
\end{figure*}

The control parameter $g_c^{(i)}$ is gradually calculated and adjusted based on the measurement results at each iteration step.
Using the Bayes update, it is possible to achieve high sensitivity~\cite{nusran2012high,cappellaro2012spin,bonato2016optimized,degen2017quantum,santagati2019magnetic,cimini2024benchmarking} and high robustness against noises~\cite{berry2002adaptive,armen2002adaptive,paesani2017experimental,wang2017experimental,han2021adaptive}  within a limited number of measurements. 
With the prior information, the initial $g_{est}^{(0)}$ can be preset as a constant.  
The value of $g_c$ for the $i$-th iteration is designed as
\begin{equation}
g_c^{(i)}=g_{est}^{(i-1)}-\pi/(2k_{\rm{eff}}T_i^2),
\label{eq:Utilizeg}
\end{equation}
which is determined by the estimated value $g_{est}^{(i-1)}$ given by the posterior distribution $p_i$ and the interrogation time $T_i$.
One can lock $P_e \approx 0.5$ to perform the optimal slope detection for each step~\cite{degen2017quantum}. 
Different from other adaptive methods that require complex calculation of auxiliary parameters~\cite{lumino2018experimental,ruster2017entanglement,PhysRevApplied.22.044058}, our design is straightforward and efficient without large amount of calculations, which is of great significance in the cases with large atom number (about $10^5 \sim 10^7$) in atomic gravimeters.

Analytically, one can find that the likelihood function is a Gaussian function versus $g$ with a standard deviation $\sigma_i = \frac{1}{\sqrt{R} k_{\rm{eff}}T_i^{2}}$, as shown in Eq. \eqref{eq:likelihoodG}. 
Then, the $i$-th posterior can be given by the multiplication of a series of Gaussian functions so that the final posterior reads
\begin{equation}\label{Gaussian_multiply}
    p_M (g|P_e; g_c^{(M)})=\frac{1}{\sqrt{2\pi} \sigma_M}\exp \left[-\frac{(g-\mu_M)^2}{2\sigma_M^2}\right],
\end{equation}
where $\mu_M \approx g$ and $\sigma_{M}=\frac{1}{\sqrt{R} k_{\rm{eff}}\sqrt{\sum_{i=1}^{M} T_i^4}}$ when $M$ is sufficiently large. 
Obviously, the standard deviation decreases as the square root of the sum of $T_i^4$ and different sequence of $T_i$ would result in different precision scaling versus the total interrogation time $\tilde{T}=\sum_{i=1}^M T_i$.
We will show more details below.

\section{Precision scaling with respect to the total interrogation time}
\noindent

For Bayesian estimation, if there is no prior knowledge or experience about the estimated parameter, it is common to set the initial prior function as a uniformly distributed function over the estimation range~\cite{lumino2018experimental}. 
Since the likelihood function for the atomic ensemble for each iteration we set up here are Gaussian functions~\cite{dinani2019bayesian}, according to Bayes' theorem, the posterior distribution can be obtained by multiplying a series of Gaussian functions.  
To evaluate the measurement precision, we focus on the standard deviations of the Gaussian functions here. 

For the first step, the result of multiplying a normalized average distribution functions by a Gaussian function is still the Gaussian function itself and the first standard deviation of posterior function 
\begin{equation}
    \sigma_1 = 1/(\sqrt{R}k_{\rm{eff}}T_1^2).
\end{equation}
While for the later steps, both the prior function and the likelihood function of the atomic ensemble are Gaussian distribution functions. When two Gaussian functions are multiplied, the result is a new Gaussian function. The normalization factor in the iterative process does not affect the standard deviation of the posterior function at each step. 
The relationship between the standard deviation ($\sigma$) of the new Gaussian function and standard deviations ($\sigma_{a}$ and $\sigma_{b}$) of the two original Gaussian distribution functions is $\sigma=\frac{\sigma_{a}\cdot\sigma_{b}}{\sqrt{\sigma_{a}^2+\sigma_{b}^2}}$.
Therefore, the standard deviation of each posterior function can be obtained from the standard deviation of the prior function and the likelihood function. Thus, we can obtain the standard deviation of the second posterior function as follows
\begin{equation}
\begin{aligned}
\sigma_2&=\frac{\frac{1}{\sqrt{R} k_{\rm{eff}}T_{1}^2}\cdot\frac{1}{\sqrt{R} k_{\rm{eff}}T_{2}^2}}{\sqrt{\left(\frac{1}{\sqrt{R} k_{\rm{eff}}T_{1}^2}\right)^{2}+\left(\frac{1}{\sqrt{R} k_{\rm{eff}}T_{2}^2}\right)^{2}}}\\
&=\frac{1}{\sqrt{R} k_{\rm{eff}}\sqrt{T_{1}^4+T_{2}^4}}.
\end{aligned}
\label{eq:sigma3}
\end{equation}
By iterating Eq. \eqref{eq:sigma3} step by step, we can obtain the $i$-th standard deviation as
\begin{equation}
    \sigma_i=\frac{1}{\sqrt{R} k_{\rm{eff}}\sqrt{\sum_{j=1}^{i} T_j^4}}.
\label{eq:sigman}
\end{equation}
Therefore, the $M$-th posterior function can be analytically calculated as 
\begin{equation}
\ p_M (g|P_e; g_c^{(M)})=\mathcal{N}\prod_{i=1}^{M}\mathcal{L}_i=\mathcal{N}\prod_{i=1}^{M}\left[\frac{1}{\sqrt{2\pi}\sigma_i}\exp(-\frac{(g-\mu_i)^2}{2\sigma_i^2})\right],
\label{eq:p_n}
\end{equation}
where $\mathcal{N}$ is the normalization factor and the standard deviation for measuring $g$ is 
\begin{equation}\label{Delta_g}
    \Delta g_{est}^{(M)}=\frac{1}{\sqrt{R} k_{\rm{eff}}\sqrt{\sum_{j=1}^{M} T_j^4}}.
\end{equation}

From Eq.~\eqref{Delta_g}, we can see how the measurements are correlated with different interrogation times in our BGE. 
If the interrogation time for each measurement is the same, for example choose $T_i=T_{max}$, the total interrogation time in the $i$-th step is $\tilde{T}_i = i T_{max}$. 
In this case we can obtain
\begin{equation}
    \Delta g_{est}^{(i)} = \frac{1}{\sqrt{R} k_{\rm{eff}}T_{max}^{3/2}\sqrt{\tilde{T}_i}} \propto \frac{1}{\tilde{T}_i^{0.5}},
\label{eq:SQL}
\end{equation} 
which is the scaling of standard quantum limit (SQL). 

If one increase $T_i$ in different manner, the precision scaling will change.  The precision scaling depends on the form of the interrogation time sequence, and it can be further improved by choosing a suitable one. 
Now we first consider the case of linearly increasing $T_{i}$. For the sake of brevity, we set $T_1=b$, where $b$ is the increment interval. Thus the $i$-th interrogation time $T_i = T_{1}+(i-1)b = ib$ and the total interrogation time in the $i$-th step is $\tilde{T}_i = i(T_{1}+T_{i})/2 \approx i^{2}b/2$ if $i \gg 1$. Substituting $T_{i}$ and $\tilde{T}_i$ into Eq. \eqref{eq:sigman}, we can obtain
\begin{equation}
\begin{aligned}
\Delta g_{est}^{(i)}&=\frac{1}{\sqrt{R} k_{\rm{eff}}\sqrt{\sum_{j=1}^i (ib)^4}}\\
&=\frac{1}{\sqrt{R} k_{\rm{eff}}\sqrt{\frac{i(i+1)(2i+1)(3i^2+3i-1)}{30}b^4}}\\
&\approx\frac{1}{\sqrt{R} k_{\rm{eff}}\sqrt{\frac{i^5}{5}b^4}}\\
&=\frac{\sqrt{5}}{\sqrt{R} k_{\rm{eff}}\tilde{T}_i^{1.25}b^{0.75}2^{1.25}} \propto \frac{1}{\tilde{T}_i^{1.25}}.
\label{eq:sigmanlinear}
\end{aligned}
\end{equation}
Therefore, we finally obtain the scaling of the standard deviation of measuring $g$ with linear increasing scheme is $\Delta g \propto \tilde{T}_i^{-1.25}$, see figure~\ref{Scaling}(a).

Then we consider the case in which $T_{i}$ grows exponentially with increment ratio $a$. Thus the $i$-th interrogation time $T_{i}=T_{1}a^{i-1}$ and the total interrogation time in the $i$-th step is $\tilde{T}_i=T_{1}(1-a^{i})/(1-a)\approx T_{i}a/(a-1)$ if $i\gg1$. Substituting $T_{i}$ and $\tilde{T}_i$ into Eq. \eqref{eq:sigman}, we can obtain
\begin{equation}
\begin{aligned}
\Delta g_{est}^{(i)}&=\frac{1}{\sqrt{R} k_{\rm{eff}}\sqrt{\sum_{j=1}^{i} T_j^4}}\\
&=\frac{1}{\sqrt{R} k_{\rm{eff}}T_{i}^2\sqrt{\sum_{j=0}^{i-1} a^{-4j}}}\\
&\approx\frac{\sqrt{a^{4}-1}}{\sqrt{R} k_{\rm{eff}}\tilde{T}_i^2(1-a)^2}\propto \frac{1}{\tilde{T}_i^{2}}.
\label{eq:sigmanunlimited}
\end{aligned}
\end{equation}
Therefore, we finally obtain the scaling of the standard deviation of measuring $g$ with exponential increasing scheme is $\Delta g \propto \tilde{T}_i^{-2}$, see figure~\ref{Scaling}(a).

Despite different precision scaling versus time exhibit for both schemes, in ideal case, the faster $T_i$ increases with larger $b$ in Eq.~\eqref{eq:sigmanlinear} or $a$ in Eq.~\eqref{eq:sigmanunlimited}, the higher measurement precision one can obtain. 

However, in practice, the length of the atomic gravimeter cavity is finite, thus $T_{i}$ cannot be increased unlimitedly.
We denote the available maximum $T_i$ as $T_{max}$ and take the exponential increasing scheme as an example. If $T_{max}$ exists, $T_i$ first grows exponentially from $T_{1}$ to $T_{max}$ and continues to stay at $T_{max}$. Assuming one needs $M_a$ steps to increase from $T_1$ to $T_{max}$ and once $T_i$ reaches $T_{max}$, it keeps fixed at $T_{max}$ for the remaining $M-M_a$ steps, i.e., 
\begin{equation}
T_i = 
\begin{cases}
   T_{max}/a^{M_a-i}, & 1 \le i <M_a,\\
   T_{max}, &  M_a \le i \le M.
\end{cases}
\label{eq:exxTi}
\end{equation}
Thus if $M_a < i \le M$, the total interrogation time in the $i$-th step is $\tilde{T}_i = T_{max}a/(a-1) + (i - M_a)T_{max} \approx (i - M_a)T_{max}$ when $i-M_a \gg {a}/{a-1}$. By substituting $T_{i}$ and $\tilde{T}_i$ into Eq.~\eqref{eq:sigman}, we can get 
\begin{equation}
\begin{aligned}
\Delta g_{est}^{(i)}&=\frac{1}{\sqrt{R} k_{\rm{eff}}\sqrt{\sum_{j=1}^{M_a} T_j^4+(i-M_a)T_{max}^4}}\\
&\approx\frac{1}{\sqrt{R} k_{\rm{eff}}\sqrt{(i-M_a)T_{max}^4}}\\
&=\frac{1}{\sqrt{R} k_{\rm{eff}}T_{max}^2\sqrt{\tilde{T}_i/T_{max}}}\\
&=\frac{1}{\sqrt{R} k_{\rm{eff}}T_{max}^{3/2}\sqrt{\tilde{T}_i}} \propto \frac{1}{\tilde{T}_i^{0.5}},
\label{eq:sigmanlimted}
\end{aligned}
\end{equation}
where the precision scaling eventually converges to the SQL $\tilde{T}_i^{-0.5}$ as Eq.~\eqref{eq:SQL} when the iteration times of using $T_{max}$ is large enough. Similarly, this result of Eq.~\eqref{eq:sigmanlimted} is also valid for using linear increasing scheme when the iteration times of using $T_{max}$ at the final stage is large enough.

\begin{figure}[htbp]
\includegraphics[width=1.0\linewidth,scale=1.00]{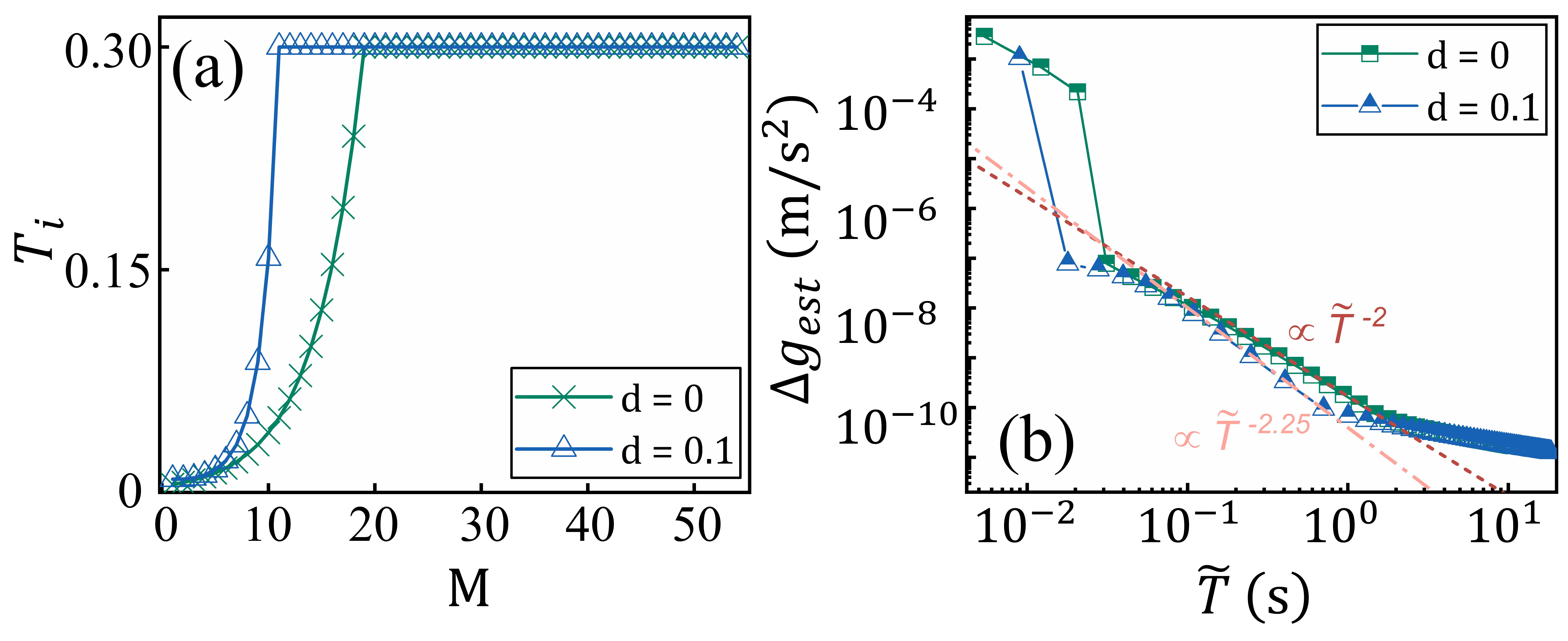}%
\caption{\label{Optimized time scheme} (a) The variation of $T_i$ according to Eq.~\eqref{eq:Ti(differenta)} with $d=0$ (green) and $d=0.1$ (blue). Here, $a_0=1.25$. (b) The gravity measurement precision $\Delta g_{est}$ versus the total interrogation time $\tilde{T}$ with $d=0$ (green) and $d=0.1$ (blue). The purple dashed and pink dash-dotted lines are the corresponding fitting curves with scaling $\propto \tilde{T}^{-2}$ and $\propto \tilde{T}^{-2.25}$, respectively. Here, $R = 5 \times 10^7$ and $k_{\rm{eff}}=1.61\times 10^7$~rad/m.}
\end{figure}

To further improve the measurement precision scaling versus interrogation time offered by our BGE, we can set a time-dependent exponential increment ratio $a(t)$, where $a(t) \rightarrow a(i)$ linear increases as the iteration number $i$ grows, as shown in Eq.~\eqref{eq:Ti(differenta)}. 
To increase the dynamic range, we also adopt the point identification method (more details are shown in Appendix C).
In this case, the interrogation time for the $i$-th iteration can be given as
\begin{equation}
T_i = 
\begin{cases}
   T_1, \quad & i \le 2 \\
   T_{i-1}\left[1.1 +(i-3)d\right], & 2< i < M_c \\
   T_{max}, & M_c \le i \le M
\end{cases}
\label{eq:Ti(differenta)}
\end{equation}
where $T_1 = T_{max} / \prod_{i=3}^{M_c} a(i)$ and $a(i)=a_0+(i-3)d$ with $a_0$ initial exponential ratio and $d$ the increment interval. 
Here, we set $a_0=1.25$. With $d=0$, it reduces to the exponential increasing scheme, see the green line in figure~\ref{Optimized time scheme}(a). While for  $d=0.1$, $T_i$ increases faster than the exponential one, see the blue line in figure~\ref{Optimized time scheme}(a).

Surprisingly, as shown in figure~\ref{Optimized time scheme}(b), the gravity measurement precision with $d=0.1$ scales $\Delta_{est}\propto \tilde{T}^{-2.25}$ versus total interrogation time $\tilde{T}$ (see the pink dash-dotted line by fitting the data), which is better than the one $\Delta_{est}\propto \tilde{T}^{-2}$ with $d=0$. 
Thus in ideal case, we find that increasing $T_i$ faster can not only improve the precision scaling versus the total interrogation time $\tilde{T}$, but also reduces the experimental measurement times needed under the same total interrogation time, which may result in higher sensitivity when $\tilde{T}$ is not large.
However, when $T_i$ reaches $T_{max}$ and the number of $T_{max}$ begins to dominate, the precision scaling will finally converge to $\tilde{T}^{-0.5}$ for both schemes, which is consistent with the analysis from Eq.~\eqref{eq:sigmanlimted}.

\begin{figure*}[htbp]
\includegraphics[width=\linewidth,scale=1.00]{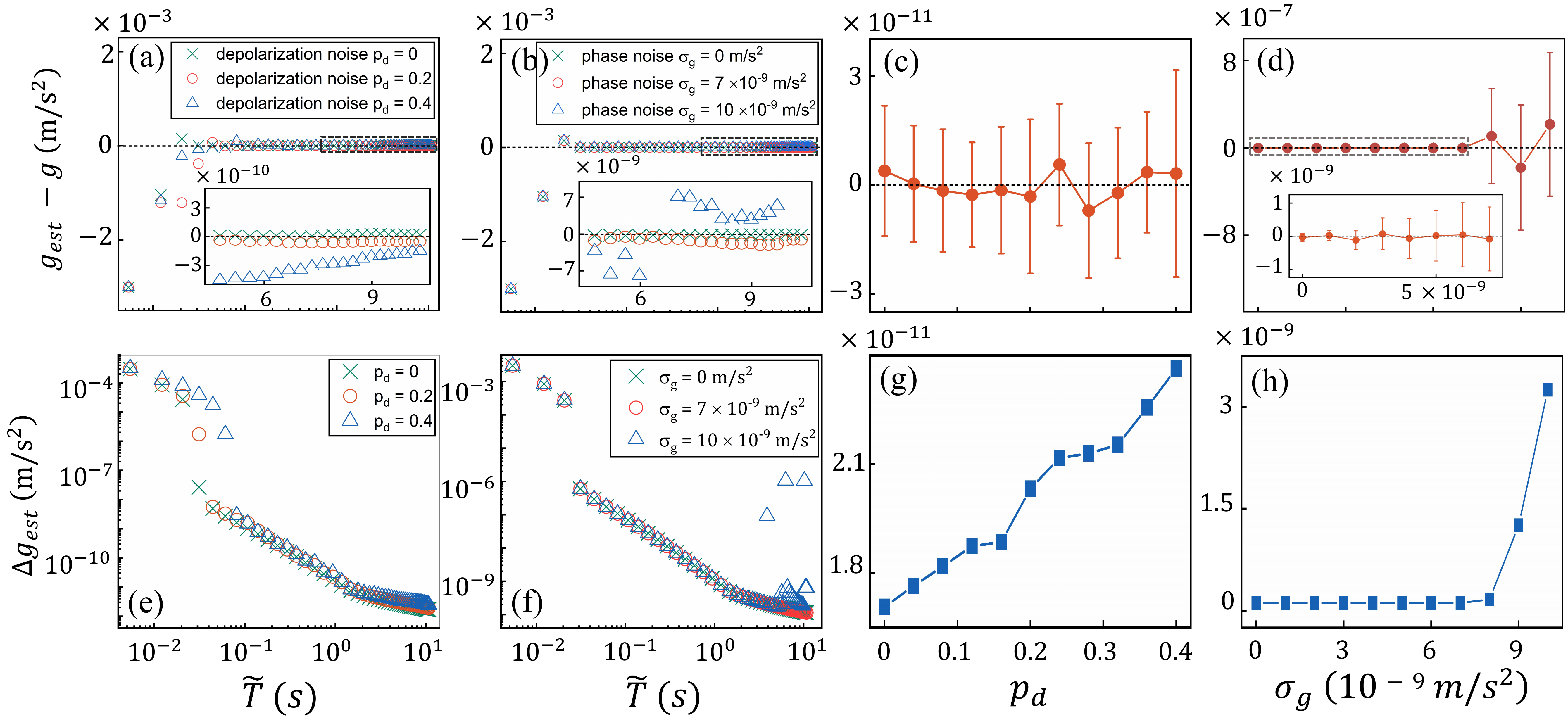}%
\caption{\label{NoiseRobust} Performance of Bayesian atomic gravimetry in the presence of noises. 
The error $g_{est}-g$ versus the total interrogation time $\tilde{T}$ for different strengths of (a) depolarization noise and (b) phase noise. 
The final error $g_{est}-g$ versus (c) the depolarization noise strength $p_d$ and (d) the phase noise strength $\sigma_g$ with error bars denoting their fluctuations for $30$ repetitions. 
The final precision $\Delta g_{est}$ versus $\tilde{T}$ for different strengths of (e) depolarization noise and (f) phase noise. 
The final precision $\Delta g_{est}$ versus (g) $p_d$ and (h) $\sigma_g$. Here, $M=50$ and $R=5\times 10^7$, $T_{max}=0.3$~s, $C=0.15$ are chosen based on typical atomic fountain experiments.}
\end{figure*}

In practise, since the interference cavity length of an atomic gravimeter is finite, there always has a limited interrogation time $T_{max}$ so that $T_i\le T_{max}$.
To design a practical sequence of $T_i$, we set the last interrogation time $T_M$=$T_{max}$ and derive the previous ones. 
If the interrogation time increases linearly, the sequence of $T_i$ can be given as 
\begin{equation}
T_i = 
\begin{cases}
   T_{max}-(M_b-i)b, & 1 < i <M_b,\\
   T_{max}, &  M_b \le i \le M.
\end{cases}
\label{eq:linearly}
\end{equation}
Given the minimum interrogation time $T_{min}$, one needs $M_b$ steps to increase from $T_1=\mathrm{max} \{ T_{min}, T_{max}-(M_b-1)b\}$ to $T_{max}$. Once $T_i$ reaches $T_{max}$, it keeps fixed at $T_{max}$ for the remaining $M-M_b$ steps. 

Similarly, if the interrogation time increases exponentially, the sequence of $T_i$ can be given as
\begin{equation}
T_i = 
\begin{cases}
   T_{max}/a^{M_a-i}, & 1 \le i <M_a,\\
   T_{max}, &  M_a \le i \le M.
\end{cases}
\label{eq:exTi}
\end{equation}
Here, $T_i$ needs $M_a$ steps to increase from $T_1=\mathrm{max} \{ T_{min}, T_{max}/a^{M_a-1}\}$ to $T_{max}$, and subsequently stays constant at $T_{max}$ for the remaining $M-M_b$ steps. 
As shown in figure~\ref{Scaling}(b) and (c), in the presence of $T_{max}$, the measurement precision scales as $\Delta g_{est} \propto \tilde{T}^{-1.25}$ for linear increasing $T_i$ and $\Delta g_{est} \propto \tilde{T}^{-2}$ for exponential increasing $T_i$. 
Subsequently, after $T_i$ reaches $T_{max}$ and remains fixed, the precision reverts to $\Delta g_{est} \propto \tilde{T}^{-0.5}$.

\section{Robustness against noises}

\noindent

One significant feature of our adaptive BGE is the robustness against noises.
The key is that our protocol uses the measurement taken with short $T_i$ for rough estimation, while relying on large quantum fluctuations in short $T_i$ interferometry to effectively mitigate the impact of noise sources.
There are two typical noises~\cite{lumino2018experimental,rambhatla2020adaptive,qiu2022efficient} in atomic gravimetry: depolarization noise and phase noise. 
Depolarization noise leads to contrast loss, while phase noise introduces random errors. 
With depolarization noise, the likelihood function becomes $\mathcal{L}_{u} = \frac{1}{2}\left[1 + (1 - \tilde p_d)(-1)^u \cos(g - g_c^{(i)}) k_{\rm{eff}}T_i^2\right]$, where $\tilde p_d \sim |\mathcal{N} (0, p_d^{2})|$ is a Gaussian distribution with $0\le p_d\le 1$. 
While in the presence of phase noise, the likelihood function becomes $\mathcal{L}_{u} = \frac{1}{2}[1 + (-1)^u C \cos(g - g_c^{(i)} + \tilde \sigma_g) k_{\rm{eff}}T_i^2]$, where the contrast $C$ is always less than 1,  and $\tilde \sigma_g \sim \mathcal{N}(0, \sigma_g^{2})$ is a Gaussian distribution with $\sigma_g \ge 0$. 

We first analyze the error $g_{est}-g$ and the measurement precision $\Delta g_{est}$ versus the total interrogation time $\tilde{T}$.
In the case of depolarization noise, the error convergence speed and the measurement precision decrease when the noise strength increases, see figure~\ref{NoiseRobust}(a) and (e).
In the case of phase noise, the errors and measurement precision almost remain unchanged for small $\sigma_g$.
However, when strong noise is present, such as $\sigma_g = 1~\mu$Gal, $g_{est}$ and $\Delta g_{est}$ will abruptly change because of the significant shift between the center positions of the likelihood function and the prior distribution.
Therefore the corresponding estimation becomes unreliable, see figure~\ref{NoiseRobust}(b) and (f).
In our calculation, we set $C=0.15$ and $R=5\times10^7$ based upon the typical atomic fountain experiments~\cite{peters1999measurement,muller2008atom,chung2009atom,hu2013demonstration}.

Then, we analyze how the measurement results depend on noise strength by averaging $30$ samples. 
In the case of depolarization noise, the errors $g_{est}-g$ and their fluctuations (denoted by errorbars) are slightly changed for small $p_d$ [see figure~\ref{NoiseRobust}(c)], and the measurement precision $\Delta g_{est}$ slightly increases with $p_d$ [see figure~\ref{NoiseRobust}(g)].
In the case of phase noise, the errors $g_{est}-g$ and their fluctuations (denoted by errorbars) are significantly changed for $\sigma_g$ [see figure~\ref{NoiseRobust}(d)], and the measurement precision $\Delta g_{est}$ is almost unchanged for small $\sigma_g$ [see figure~\ref{NoiseRobust}(h)].
However, errors and measurement precision may deteriorate, leading to unreliable results when there is significant phase noise, as observed in cases where $\sigma_g \geq 0.9~\mu$Gal.

Phase noise and depolarization noise may be present simultaneously in the interference process of atomic gravimeter. 
When the BGE adaptively determines the linear chirp rate of the Raman laser, these two types of noises may have an influence on obtaining the optimal measurement results.
The phase noise of the gravimeter is mainly due to the phase noise of the Raman laser and the vibration noise of the mirror.
It is considered as random noise of a Gaussian distribution, which causes the phase to be measured to deviate from the actual value.
We have investigated the influences of phase noise, and here we further consider the influence of contrast. 

\begin{figure}[htbp]
\includegraphics[width=0.6\linewidth,scale=1.00]{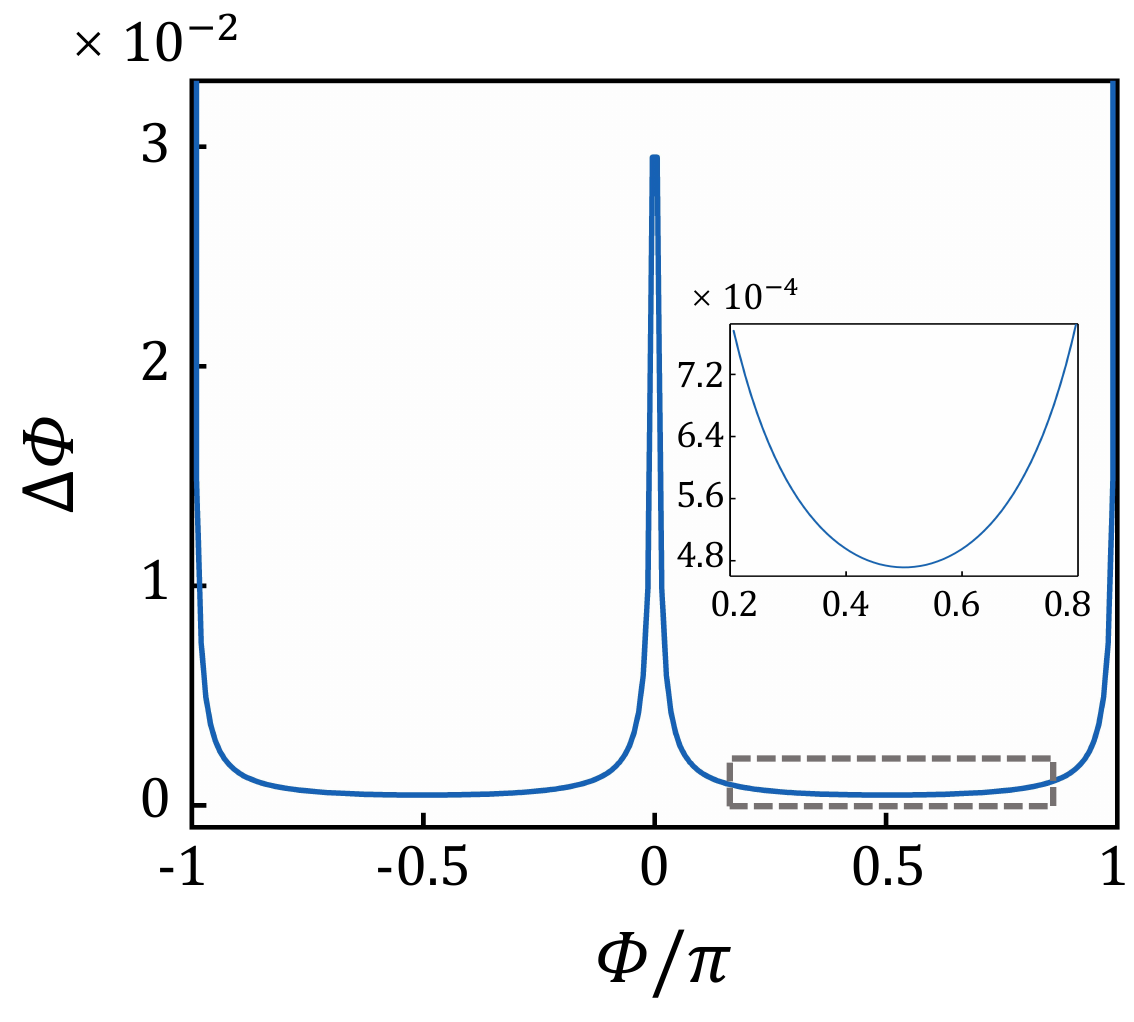}%
\caption{\label{Error propagation formula} The phase estimation precision under the noises when $C=0.15$ in the phase range $[-\pi, \pi]$, while the inset is the enlarged area in the phase range $[0.2\pi, 0.8\pi]$}
\end{figure}

Depolarization noise mainly caused by the thermal velocity distribution of the atomic cloud and the limited size of the laser beam result in atoms at different positions within the same light pulse not achieving exactly the same transition probability, which would lead to a reduced contrast $C<1$ in 
\begin{equation}
    P_u= \frac{1}{2}\left[1+(-1)^u C\cos\Phi\right],
\end{equation}
where $u=0$ or $1$ represent the atom occupies $\ket{g, \textbf{p}}$ or $\ket{e, \textbf{p}+ \hbar\textbf{k}_{\rm{eff}}}$, respectively~\cite{samuel2022using}.
It may also arise when the $\pi$ and $\pi/2$
pulses do not act as perfect mirrors or beam-splitters. 
Assuming that the thermal velocity distribution of the atomic cloud and the distribution and intensity of the laser are the same for each experiment, the contrast is close for each experiment.
Therefore, it can be assumed that under the same experimental condition, the contrast $C$ is a constant for each measurement.

Considering a fixed contrast $C$, according to error propagation formula, we can obtain the phase estimation precision over a period

\begin{equation}
\begin{aligned}
\Delta \Phi&=\frac{\sigma_p}{|\partial P/\partial \phi|}\\
&=\frac{\sqrt{\frac{1}{R}\frac{1}{2}\left[1-C\cos(\Phi)\right]\frac{1}{2}\left[1+C\cos(\Phi)\right]}}{|\frac{C}{2}\sin(\Phi)|}
\label{eq:errorpropagation}
\end{aligned}
\end{equation}
Here, $\sigma_p$ denotes the measurement fluctuations in probability and we assume $\sigma_p=1/\sqrt{R}$ with $R$ the atom number.
Ideally with $C=1$, one can easily return to the case with $\Delta \Phi = \frac{1}{\sqrt{R}}$, which is independent on The value of $\Phi$ itself. 
While in the presence of depolarization noise, the contrast is reduced with $C < 1$, and different phase value would result in different phase estimation precision, as shown in figure~\ref{Error propagation formula}.
In this case, one can find that $\Phi=\pm\pi/2$ corresponds to the optimal measurement with the highest measurement precision, see the inset in figure~\ref{Error propagation formula}. 
For one thing, $\Phi=\pi/2$ is not only the most sensitive measurement point~\cite{degen2017quantum} with $\Delta \Phi = 1/(\sqrt{R} C)$, but also corresponding to the highest measurement precision, so BGE locks $\Phi=\pi/2$ for each measurement to get the most sensitive and precise measurement.
Therefore we can get $\Delta g=\frac{1}{C\sqrt{R}k_{\rm{eff}}T_i^2}$ for each measurement and we choose to lock $\Phi=\pi/2$ for each iteration in our BGE.

\section{Precision improvement of realistic atomic gravimeters}
\noindent

\begin{figure}[htbp]
\includegraphics[width=1\linewidth,scale=1.00]{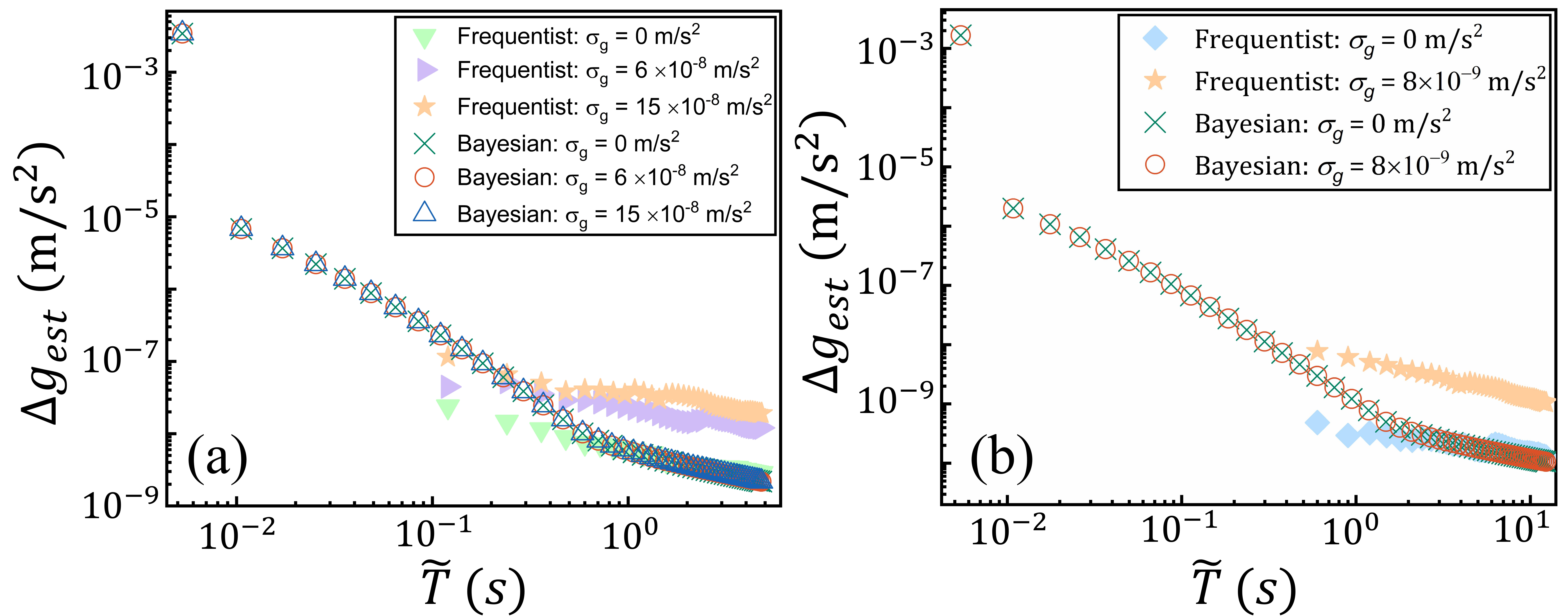}%
\caption{\label{Comparison} (a) Improvement of a transportable gravimeter under phase noise via our Bayesian protocol. Here, we use an exponential increasing scheme with $a = 1.25$, $T_{max} = 0.12$~s, $C = 0.16$,  $R = 5 \times 10^6$, and $k_{\rm{eff}} = 1.47\times 10^{7}$~rad/m as in ~\cite{wu2019gravity}. (b) Improvement of a Rb fountain gravimeter under phase noise via our Bayesian protocol. Here, we use an exponential increasing scheme with $a = 1.25$, $T_{max} = 0.3$~s, $\tilde{T} = 12$~s, $C = 0.15$, $R = 5 \times 10^7$, $k_{\rm{eff}} = 1.61\times 10^{7}$~rad/m. as in~\cite{hu2013demonstration}.}
\end{figure}

In conventional frequentist estimation, one may choose $T_{max}$ to achieve the highest precision, but the dynamic range becomes smallest due to phase ambiguities.
In our BGE, one may choose a suitable minimum interrogation time $T_{min}$ to ensure a high dynamic range, whereas increasing $T_i$ improves the measurement precision.
Moreover there is no need to scan at least three different fringes to pre-estimate $g$.
Notably, the measurement precision can be significantly enhanced under noisy condition, opening up promising practical applications.

In our simulation, we consider a transportable atomic gravimeter with Cs atoms~\cite{wu2019gravity} as an example.
Fixing the total interrogation time as $\tilde{T}=40~T_{max}$, we assume the conventional frequentist protocol takes $40$ measurements with $T_{max}=0.12~s$, while our BGE performs $51$ measurements with $a=1.25$. 
Figure~\ref{Comparison}(a) shows the measurement precisions versus the interrogation time $\tilde{T}$.
Although the precisions are almost the same when $\sigma_g=0$, the improvement becomes significant when noises appear.
The precision of the conventional protocol rapidly decreases with the noise strength.
When $\sigma_g = 6\times 10^{-8}$~m/s$^2$, our BGE can achieve a precision that is $5$ times greater than that of the conventional method.
Moreover, the stronger the phase noise, the more significant this improvement will be. 
While for the state-of-the-art fountain atomic gravimeter and compare the measurement precision achieved by our BGE and conventional method under different experimental conditions. Here, we set $k_{\rm{eff}} = 1.61\times 10^{7}$~rad/m, $T_{max} = 0.3$~s, $C = 0.15$, and $R = 5 \times 10^7$ as achieved in experiment~\cite{hu2013demonstration}.
Assuming the conventional method takes $40$ measurements in total to perform the gravity measurement, the Bayesian method will apply $56$ measurements for the same total interrogation time in the case of $a=1.25$. 
When $\sigma_g=0$, the final precision is almost the same of two method. 
The improvement becomes significant when noises appear.
The precision achievable through the conventional method rapidly decreases with the noise strength.
For a Rb atomic gravimeter under the phase noise with $\sigma_g = 8\times 10^{-9}$~m/s$^2$, our BGE can achieve a precision that is 10 times greater than that of the conventional method, see figure~\ref{Comparison}(b).


Increasing the interrogation time can improve the measurement precision, the current maximum interrogation time that can be applied to gravimeter is 1.15 s~\cite{dickerson2013multiaxis}.
However, a larger interrogation time means a narrower Gaussian distribution of the likelihood function, and a greater impact of phase noise, resulting in faster unreliable result in the BGE iteration process. 
Meanwhile, the increase in interrogation time can also cause several problems.
The first is that the length of the interference cavity of the gravimeter increases dramatically. 
The second is the need to reduce the temperature of the atomic cloud to the order of tens of nK, in order to reduce the effects of decoherence over such a long interrogation time. 
Finally, the longer the interrogation time, the narrower the Gaussian distribution of the likelihood function, and the greater the phase noise introduced, which makes the faster unreliable result in the BGE iteration.

\begin{figure}[htbp]
\includegraphics[width=\linewidth,scale=1.00]{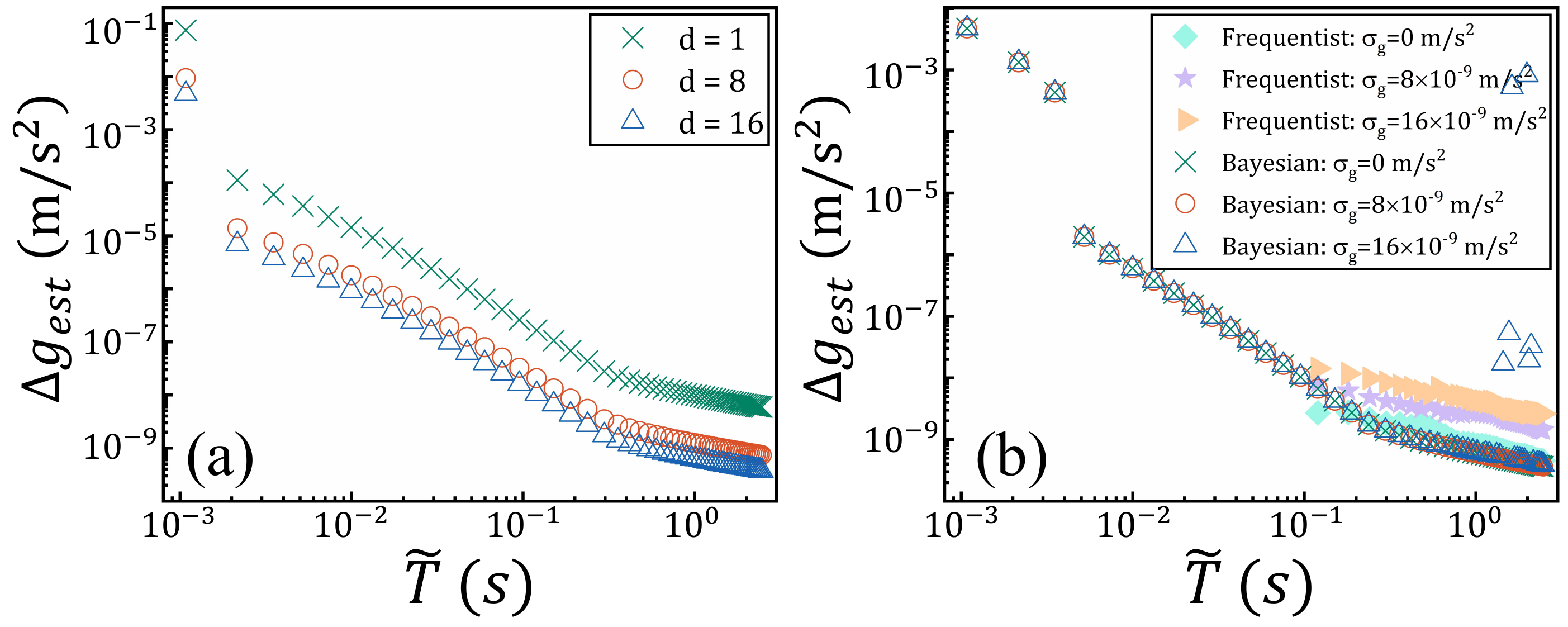}%
\caption{\label{Bragg comparision} Performances of BGE and conventional methods in Bragg atomic gravimeters. (a) The measurement precision of gravity using BGE with different diffraction order $d$. (b) Comparison of the precision with BGE and conventional frequentist method under different phase noise strength $\sigma_g$ with diffraction order $d = 16$. Here, we set $T_{max} = 60$~ms, $a = 1.25$, $R = 1 \times 10^7$, and $C = 0.15$.}
\end{figure}

To improve the precision of quantum gravimeters, one can also try to the interferometric area to make the effective momentum as large as possible~\cite{altin2013precision,mazzoni2015large,abend2016atom,hardman2016simultaneous,cheng2018momentum,zhang2023ultrahigh}. 
It has been demonstrated that one can increase the effective momentum via large momentum transfer using Bragg diffraction. 
This allows to improve the sensitivity by increasing the effective separation between the two interfering atomic samples.
Thus applying our BGE algorithm to the Bragg atomic gravimeters may improve the gravity measurement precision even further. 
Unlike the Raman atomic gravimeter, the phase accumulation of the Bragg atomic gravimeter becomes 
\begin{equation}
    \Phi = d (k_{\rm{eff}}g-2\pi \alpha) T_{i}^2,
\end{equation}
where $d$ means the diffraction order and $T_i$ is the interrogation time for the $i$-th Bayesian iteration. 
It obtains the transfer of multiphoton momentum without changing the internal state of the atom, which can reduce the influence of electromagnetic field fluctuations on the probability of internal state transition. 
Most importantly, it can offer $d$ times enhancement under the same interrogation time due to $d$ order of Bragg transition.
As shown in figure~\ref{Bragg comparision}(a), the precision obtained by BGE is proportional to the diffraction order $d$, which means the higher the diffraction order, the higher measurement precision can be.
%
%
We also show the performances of BGE in comparison with conventional frequentist method under phase noise.
On one hand, the presence of phase noise makes the transition probability deviate from the noise-free situation, resulting in large fluctuations in the fringe fitting method.
The effect of $\sigma_g$ related to phase noise is more pronounced at the higher the diffraction order of the Bragg atomic gravimeter, and in figure~\ref{Bragg comparision}(b) we show that the effect of noise is at a diffraction order of $16$.
On the another hand, BGE still achieves high precision in the same phase noise strength.
When the $\sigma_g$ is less than $8\times 10^{-9} \rm{m/s^2}$, the phase noise has little effect on the precision. 
However, as $\sigma_g$ continues to increase, the center of the likelihood function and the prior function deviate too much, which will cause iterative errors and large fluctuations in precision, see the triangles in figure~\ref{Bragg comparision}(b).

\section{Summary and discussions}
\noindent
We propose an adaptive Bayesian quantum estimation protocol to achieve robust high-precision gravimetry. 
Based upon a tailored sequence of correlated Mach-Zehnder interferometry taken with short to long interrogation times, our protocol allows us to incorporate prior knowledge of gravity and update it with Bayes' theorem to obtain high precision.
Under noisy conditions, it is possible to improve the measurement precisions of state-of-the-art atomic gravimeters by an order of magnitude without compromising the dynamic range. 
Our protocol not only has promising application for versatile gravity measurement necessaries such as transportable gravimeters~\cite{bidel2018absolute,bongs2019taking,Zhang2020, Xu2022,zhong2022quantum,chen2023analysis} for exploring mineral and oil resources, but also applicable in various interferometry-based quantum sensors such as atomic clocks, quantum magnetometers, and atomic gyroscopes.

\section*{Data availability statement}

\noindent
Data that support the figures within this paper and other findings of this study are available from the corresponding authors upon reasonable request.

\section*{Acknowledgments}
\noindent
Jinye Wei and Jiahao Huang contributed equally to this work. This work is supported by the National Natural Science Foundation of China (Grants No.~12025509, No.~12475029 and No. 92476201), the National Key Research and Development Program of China (Grant No.~2022YFA1404104), and the Guangdong Provincial Quantum Science Strategic
Initiative (GDZX2305006 and GDZX2405002).

\appendix

\section*{Appendix A: Principle of an atomic gravimeter}\label{appendixa}

\noindent
An atomic Mach-Zehnder interferometry, which employs a $\pi/2-\pi-\pi/2$ Raman pulse sequence to coherently split, reflect, and combine the atomic wave packets, is widely used for measuring the gravitational acceleration $g$. 
The beam-splitters ($\pi/2$ pulses) and mirrors ($\pi$ pulses) are achieved by state-changing Raman transitions, which are achieved with two counter-propagating laser pulses coherently coupling two internal states $\ket{g}$ and $\ket{e}$. 
The atomic gravimeter uses two laser beams with wavevector $\textbf{k}_1$ and $\textbf{k}_2$ that enter from opposite directions to couple the atomic transition between $\ket{g}$ and $\ket{e}$.

The free-falling and fountain geometries are often used in atomic gravimeters.
For the atomic free-falling gravimeter, the atoms are cooled above the interference zone and released to fall freely under the action of gravity~\cite{merlet2010comparison,carraz2009compact,altin2013precision,chen2020portable}. 
While for the atomic fountain gravimeter, one first cools the atoms in the lower part of the interference zone, adjusts the detuning of the upper and lower lasers, so that the atomic cluster obtains a vertical upward speed to throw up, and then fall freely~\cite{peters2001high,chung2009atom,zhou2011measurement,freier2016mobile}. 
With the same interference region length, the maximum interrogation time of the atomic fountain gravimeter can be twice that of the atomic free-fall gravimeter, which can achieve higher measurement precision.

Commonly for an atomic gravimeter, all atoms are prepared in $\ket{g, \textbf{p}}$ at the beginning. 
Here, $\textbf{p}$ is labelled as the momentum.
In order to generate the spatial separation needed for gravimetry,  the Raman pulse transits the atoms from $\ket{g, \textbf{p}}$ to $\ket{e, \textbf{p}+ \hbar\textbf{k}_{\rm{eff}}}$, obtaining an additional effective momentum to separate the two states, where $\hbar \textbf{k}_{\rm{eff}}=\hbar (\textbf{k}_1-\textbf{k}_2)$ and $\textbf{k}_{1,2}$ are the wavevectors of the two Raman lasers.
To make the effective momentum largest, the two lasers are counter-propagating along the gravity direction $\hat z$, and $\textbf{k}_{1}=k\hat{z}$, $\textbf{k}_{2}=-k\hat{z}$.
In this two-photon process, the photon with momentum $\hbar \textbf{k}_1$ is incident from the lower part of the atom, while the photon with momentum $\hbar\textbf{k}_2$ is incident in the opposite direction, resulting in an effective momentum $\hbar \textbf{k}_{\rm{eff}}=\hbar k_{\rm{eff}} \hat {\textbf{z}}=\hbar (\textbf{k}_1-\textbf{k}_2)= 2\hbar \textbf{k}=2\hbar k \hat {\textbf{z}}$.
The first Raman $\pi/2$ pulse places the atoms in an equal superposition of states $\ket{g, \textbf{p}}$ and $\ket{e, \textbf{p}+ \hbar\textbf{k}_{\rm{eff}}}$, and the atoms evolve freely with interrogation time $T$.
Then, another Raman $\pi$ pulse completely exchanges the two atomic states (i.e., $\ket{g, \textbf{p}} \rightleftarrows \ket{e, \textbf{p}+ \hbar\textbf{k}_{\rm{eff}}}$) and reflects the two paths of atoms as mirrors. 
After another free evolution with the same interrogation time $T$, the second Raman $\pi/2$ pulse is applied for recombination and one can perform the population detection to extract the gravity.

Mathematically, the Raman beam splitters can be described by a $2\times2$ matrix~\cite{kritsotakis2018optimal}
\begin{equation}
U(t,\tau,\phi) = \begin{bmatrix}
\cos(\frac{\Omega \tau}{2})e^{-i\omega_g\tau} & -ie^{-i (k_{\rm{eff}}\hat{z}-{\phi})}\sin(\frac{\Omega \tau}{2})e^{-i\omega_g\tau} \\
-i e^{i(k_{\rm{eff}}\hat{z}-{\phi})} \sin(\frac{\Omega \tau}{2})e^{-i\omega_e\tau} & \cos(\frac{\Omega \tau}{2})e^{-i\omega_e\tau}
\end{bmatrix},
\label{eq:operateru}
\end{equation}
where $\Omega$, $\tau$ and $\phi$ denote the Rabi frequency of the two-photon Raman transition, the duration of evolution and the phase of Raman laser, respectively. In addition, ${\Phi}$ denotes the initial phase of laser and $\hat{z} = v_{0}t - g t^2/2$ stands for the instant location of the atomic cloud with $t$ the beginning time of evolution.
The $\pi/2$-pulse and $\pi$ pulse are achieved with $\Omega\tau = \pi/2$ and $\Omega\tau = \pi$, respectively.

Suppose the Raman beam splitters acting on the atoms are instantaneous processes, the $\pi/2$-pulse can be respectively written as
\begin{equation}
U(t,\tau/2,\phi) = \begin{bmatrix}
\frac{\sqrt{2}}{2} & -ie^{-i (k_{\rm{eff}}\hat{z}-{\phi})}\frac{\sqrt{2}}{2} \\
-i e^{i(k_{\rm{eff}}\hat{z}-{\phi})} \frac{\sqrt{2}}{2} & \frac{\sqrt{2}}{2}
\end{bmatrix},
\end{equation}
and the $\pi$-pulse can be respectively written as
\begin{equation}
U(t,\tau,\phi) = \begin{bmatrix}
0 & -ie^{-i(k_{\rm{eff}}\hat{z}-{\phi})} \\
-i e^{i(k_{\rm{eff}}\hat{z}-{\phi})}  & 0
\end{bmatrix}.
\end{equation}

While for the free evolution process during interrogation time $T$, the corresponding matrix is only related to the eigenenergies of the two levels $\omega_{g,e}$, which is in the form of
\begin{equation}
U(t,T,0) = \begin{bmatrix}
e^{-i\omega_{g} T} & 0 \\
0 & e^{-i\omega_{e} T}
\end{bmatrix}.
\label{eq:operaterfreeevolution}
\end{equation}

The whole procedure of the evolution can be reached by the product of three Raman transport matrices and two free evolution matrices, which can be respectively written as
\begin{equation}
U_{total} = U(t_3,\tau/2, \phi_3)U(t_2+\tau,T, 0)U(t_2,\tau, \phi_2)U(t_1+\tau/2,T, 0)U(t_1,\tau/2, \phi_1)
\label{eq:operaterwholeevolution}
\end{equation}
Assuming all atoms occupy $\ket{g, \textbf{p}}$ initially, after the whole interference process, the final state reads
\begin{equation}
\begin{aligned}
\ket{\psi}_f&=U_{total}\ket{g,\textbf{p}}
    \\
    &=-\frac{1}{2}e^{i(\omega_e+\omega_g)T}\left(
\left[e^{i(\phi_1-\phi_2)}+e^{i(\phi_2-\phi_3)}\right]\ket{g,\textbf{p}} 
+\left[ie^{i\phi_2}-ie^{i(\phi_1-\phi_2+\phi_3)}\right]\ket{e, \textbf{p}+ \hbar\textbf{k}_{\rm{eff}}}\right).
\label{eq:finalstate}
\end{aligned}
\end{equation}
Here, $\phi_i=k_{\rm{eff}}\hat{z}_i-\phi$ represent the phases of the first $\pi/2$ pulse, $\pi$ pulse and the second $\pi/2$ pulse with $\hat{z}_i=v_0t_i-gt_i^2/2$ ($i=1,~2,~3$).
Due to the symmetrical evolutionary geometry of the gravimetry and assume the duration of pulses is negligible, the evolution time satisfy $(t_1+t_3)/2=t_2$ and $t_3=t_2+T=t_1+2T$.
Thus the normalized signal for detecting atoms in the two states are~\cite{baryshev2015application}  
\begin{equation}
    \mathcal{L}_{u} = \frac{1}{2}\left[1 + (-1)^u \cos(k_{\rm{eff}}g T^2)\right],
\label{eq:singlelikelihood}
\end{equation}
where $u = 0$ or $1$ stand for the atom occupying $\ket{g, \textbf{p}}$ or $\ket{e, \textbf{p}+ \hbar\textbf{k}_{\rm{eff}}}$, respectively. 

In order to cancel out the phase ambiguity to get $g$ accurately, conventional atomic gravimeters need to scan $\alpha$ to obtain the interferometry fringes with likelihood function being 
\begin{equation}
    \mathcal{L}_{u} = \frac{1}{2}\left[1 + (-1)^u \cos(k_{\rm{eff}} g-2\pi \alpha) T^2\right].
\label{eq:slikelihood}
\end{equation}
Here, $\alpha$ denotes the chirp rate of the Raman beams.
It is necessary to scan at least three fringes with different interrogation time $T$ to obtain a common $\alpha_0$ determining $g=2\pi \alpha_0 / {k}_{\rm{eff}}$ in the pre-estimation stage~\cite{carraz2009compact,zhou2011measurement,abend2016atom,freier2016mobile,samuel2022using}.
Once $\alpha_0$ is determined, the gravimeters will work with interrogation time $T_{max}$ as long as possible.

\section*{Appendix B: Procedure of Bayesian gravity estimation} \label{appendixb}

\begin{algorithm}[H]
    \SetKwInOut{Input}{Input}
    \SetKwInOut{Output}{Output}
    \SetKwInOut{Initialize}{Initialize}
    \SetAlgoLined

    \Input{total atom number $R$; maximum interrogation time $T_{\text{max}}$; exponential increasing rate $a$; }

    \Initialize{minimum interrogation time $T_1$; initial interval $[g_l^{(1)}, g_r^{(1)}]$; initial prior uniform distribution $p_{0}(g)=1/(g_r^{(1)}-g_l^{(1)})$}

    {[Main loop]}
    
    \For{$i=1$ \KwTo $M$}{
        [Updates of parameter]\;
        
        $T_i = 
        \begin{cases}
        T_{max}/a^{M_a-i}, & 1 \le i <M_a\\
        T_{max}, & M_a \le i \le M

        \end{cases}$
        
        
   
        Length of the interval: $g_{lr}^{(i)} = g_{r}^{(i)}-g_{l}^{(i)}=\frac{2\pi}{k_{\text{\rm{eff}}}T_i^2}$\;
        
        \If{$T_i \neq T_{i-1}$}{
        $g_l^{(i)} \leftarrow g_{est}^{(i)} - {g_{lr}^{(i)}}/{2}$\;
        
        $g_r^{(i)} \leftarrow g_{est}^{(i)} + {g_{lr}^{(i)}}/{2}$\;
        
        Reset the prior function: $p_{i-1}(g) = \frac{1}{\sqrt{2\pi}\sigma_i}\exp\left[\frac{(g - \mu_i)^2}{2 \sigma_i^2}\right]$, where $\mu_i = g_{est}^{(i)}$ and $\sigma_i = \Delta g_{est}^{(i)}$\;
        }
        Linear chirp rate of Raman laser: $\alpha_c^{(i)} = g_c^{(i)} k_{\rm{eff}}/2\pi$ and $g_c^{(i)}\leftarrow g_{est}^{(i-1)}-\pi/(2k_{\rm{eff}}T_i^2)$\;
        \BlankLine
        [Experimental measurement]\;
        measured population signal $P_e$ using $T_i$ and $\alpha_c^{(i)}$\;
        \BlankLine
        [Bayesian iteration]\;
        
        Likelihood function: $\mathcal{L}(P_e|g; g_c^{(i)}) = \frac{1}{\sqrt{2\pi} \sigma}\exp\left[-\frac{(P_e - \mathcal{L}_{u}(1|g; g_c^{(i)}))^2}{2\sigma^2}\right]$, where $\mathcal{L}_{u}(1|g; g_c^{(i)}))=\frac{1}{2}\left\{ 1 - \cos [(g-g_c^{(i)}) k_{\rm{eff}} T_i^2]\right\}$ and $\sigma^2 \approx P_e(1-P_e)/R$\;

        Bayesian update: $p_i(g|P_e; g_c^{(i)}) \leftarrow \mathcal{N}\mathcal{L}(P_e|g; g_c^{(i)})p_{i-1}(g)$\;

        mean of the estimator: $g_{est}^{(i)} = \int gp_i(g|P_e; g_c^{(i)}) dg$\;
        
        standard deviation of the estimator: $\Delta g_{est}^{(i)} = \sqrt{\int g^2p_i(g|P_e;   g_c^{(i)})\, dg - (g_{est}^{(i)})^2}$ \;
        \BlankLine
        \Output{mean of the estimator $g_{est}^{(i)}$; standard deviation of the estimator $\Delta g_{est}^{(i)}$\;}
    } 
        
    \caption{Flowing chart of Bayesian gravity estimation}   \label{chart}    
\end{algorithm}

\section*{Appendix C: Point identification method}\label{appendixc}
\begin{figure*}[htp]
\includegraphics[width=0.7\linewidth,scale=1.00]{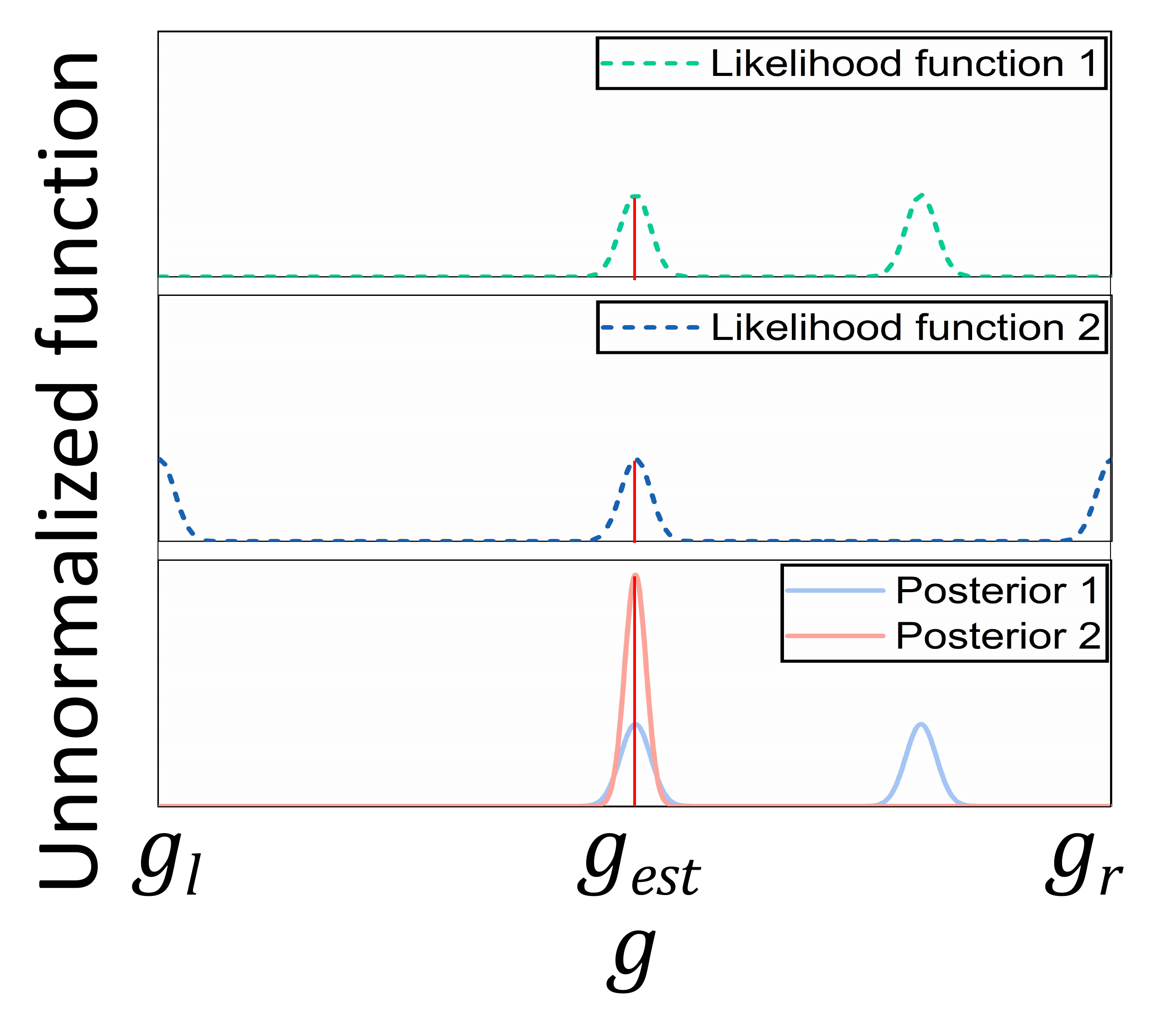}%
\caption{\label{Point identification} The schematic graph of using point identification method. Here, we assume $g_{est}$ is at the center and $g_l$ and $g_r$ denote the minimum and maximum values of the parameter estimation range for gravitational acceleration $g$, respectively. The schematics of likelihood and posterior for the first and second steps are shown. }
\end{figure*}
For atomic gravimeters, the effective wave vector of the Raman laser $k_{\rm{eff}}\sim 10^7 $~rad/m and the total atom number $R \sim 10^7$ result in the linewidth of the Gaussian-shaped likelihood function to be of the order of $10^{-7}$ or even lower. 
In this case, if the interrogation time $T_i$ changes too dramatically in the first few steps, the BGE may lock the wrong neighbouring peak of the initial likelihood function, which results in a half-period difference between $g_{est}$ and the true value $g$. 

In order to eliminate this mistake, we adopt the point identification method~\cite{PhysRevA.109.042412}, see figure \ref{Point identification}. We can set the first and second interrogation times to be equal $T_2=T_1$ with different $g_c^{(1)}$ and $g_c^{(2)}$.
With the same $T_1$ but different $g_c$, the first and second likelihood functions have only one common peak ensuring the second posterior has only one peak, which can avoid the mistake for determining the central peak and thus improve the dynamic range of our BGE.
In this case, the time sequence for $T_i$ in the case of exponential increasing used in the main text becomes

\begin{equation}
T_i = 
\begin{cases}
   T_{max}/a^{M_a-1}, &~i \leq 2\\
   T_{max}/a^{(M_a-i+1)}, &~ 2 < i < M_a\\
   T_{max}, & M_a \le i \le M+1
\end{cases}
\label{eq:n=2}
\end{equation}
where only one additional measurement with $T_1$ is used in the second step with a different $g_c$. 

\section*{Appendix D: Influence of atom number on measurement precision}\label{appendixd}

The number of atoms used in the gravimeters also affects the measurement precision and it is different with different experimental conditions.
For example, the number of atoms is generally on the order of $10^5$ to $10^7$.
In order to analyze the influence of atom number, we calculated with three different atom number $R = 5 \times 10^3, \ 5 \times 10^5, \ 5 \times 10^7$ under the contrast of $C = 0.15, 1$, respectively.
The fluctuation of the gravity measurement is inversely proportional to $\sqrt{R}$, which means that more atoms results in higher measurement precision of gravity~\cite{samuel2022using}.

\begin{figure}[htbp]
\includegraphics[width=0.75\linewidth,scale=1.00]{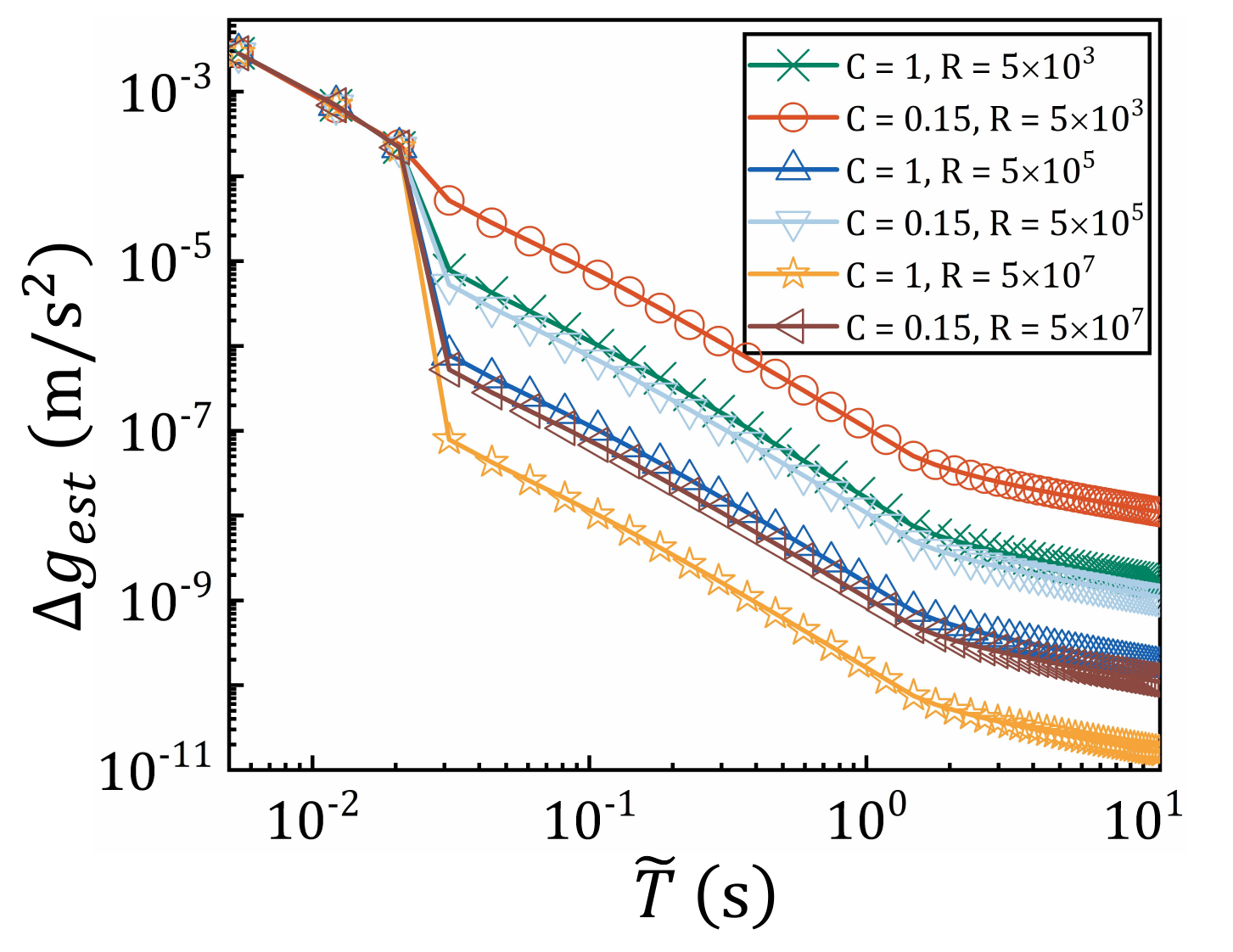}%
\caption{\label{Experimental feasibility} The influence of atom number $R$ on measurement precision with different contrast $C$.}
\end{figure}

According to Eq.~\eqref{eq:errorpropagation} with $\Phi=\pi/2$, for the same contrast $C$, the ratio between the measurement precision with atom number $R_1$ and the one with atom number $R_2$  satisfies the following relation $ \frac{\Delta g_{est}(R_1)}{\Delta g_{est}(R_2)}=\frac{\sigma_{R_1}}{\sigma_{R_2}}= \frac{\sqrt{R_2}}{\sqrt{R_1}}$, which can be verified by numerical results in figure~\ref{Experimental feasibility}.
Thus in practice one may also use the above relation to calculate the measurement precision for large $R_1$ from small $R_2$, which can improve the calculation efficiency of our BGE.

The atomic gravimeter can improve the precision by increasing the number of trapped atoms and improving the contrast, which is still a challenge in the current experiments. 
In addition, the use of quantum entanglement is also an effective way to improve the precision of gravity measurement~\cite{10.1063/5.0204102}.
However, as far as the current experimental conditions are concerned, for the atomic gravimeter with a large atomic number, the more suitable entangled state is the spin squeezed state, but the experimental system is more demanding, which is also a problem that needs to be solved urgently.

\end{document}